\documentstyle[twoside,fleqn,mod,graphicx,psfrag]{article}

\newcommand{\bea}{\begin{eqnarray}}
\newcommand{\eea}{\end{eqnarray}}
\newcommand{\simgt}{\hbox{ \raise3pt\hbox to 0pt{$>$}\raise-3pt\hbox{$\sim$} }}
\newcommand{\simlt}{\hbox{ \raise3pt\hbox to 0pt{$<$}\raise-3pt\hbox{$\sim$} }}

\title{Understanding Heavy Quarkonium Systems
       in Perturbative QCD\thanks{
Talk given at ``International Conference in Quantum Chromodynamics
(QCD02)'', Montpellier, France, July~2--9, 2002.}\thanks{
TU--667, September 2002.
} 
}

\author{Y.~Sumino\address{Department of Physics, Tohoku University,
Sendai, 980-8578 Japan}
}

\begin{document}

\begin{abstract}
\noindent
We review the recent theoretical progress 
in heavy quarkonium spectroscopy 
within the boundstate theory based on perturbative QCD.
New microscopic pictures of the heavy quarkonium systems
are obtained.
\vspace{-3mm}
\end{abstract}

\maketitle

\section{Introduction}

Recent advent of the boundstate theory based on perturbative QCD enabled
accurate descriptions of the
nature of the heavy quarkonium systems
such as bottomonium states.
Through the progress became more and more clear the importance to
eliminate properly contributions of infrared (IR) degrees of freedom
in the theoretical descriptions of these systems.
Indeed these systems comprise natural IR cutoffs both of 
{\it spacial} and {\it temporal} dimensions.
As a result, once we decouple the IR
contributions properly, we find much better convergence of
perturbative expansions and a good control over the theoretical predictions.
In order to realize IR decoupling in a systematic way,
the language of renormalons and their cancellations have played
central roles:
uncertainties of the
perturbative expansions are estimated from their asymptotic behaviors 
based on renormalon dominance picture.

In the heavy quarkonium systems, we may classify the mechanisms of IR
decoupling into two categories:
(I) The spacial size of a quarkonium state
acts as an IR cutoff.  
This leads to cancellation of ${\cal O}(\Lambda)$ renormalons, with residual
${\cal O}(\Lambda^3)$ renormalons.
(II) Offshellness of the heavy quark and antiquark acts as an IR cutoff in
temporal dimension.
This leads to cancellation of ${\cal O}(\Lambda^3)$ renormalons, 
with residual ${\cal O}(\Lambda^4)$ renormalons.
These aspects point to new and detailed
physical pictures of the heavy quarkonium
states.

\section{Cancellation of {\boldmath ${\cal O}(\Lambda)$} Renormalons}

Cancellation of ${\cal O}(\Lambda)$ renormalons
\cite{renormalon}
is a realization of the fact that the system under consideration
is color-singlet and has a spacial size $R$ much smaller than
the typical hadron size $\Lambda R \ll 1$.

\subsection{Total energy of a static $b\bar{b}$ pair}

Let us first neglect the kinetic energy of $b$ and $\bar{b}$
and examine the total energy of a static $b\bar{b}$ pair.
It is defined as the sum of the static QCD potential
and the pole masses of $b$ and $\bar{b}$:
\begin{eqnarray}
E_{\rm tot}(r) = 2 m_{b,{\rm pole}} + V_{\rm QCD}(r) .
\label{totene}
\end{eqnarray}
The ${\cal O}(\Lambda)$ renormalons contained in the pole mass and
the QCD potential are cancelled if we rewrite the pole mass
in terms of the $\overline{\rm MS}$ mass,
$\overline{m}_b \equiv m_b^{\overline{\rm MS}} (m_b^{\overline{\rm MS}})$.
We thus substitute
$m_{b,{\rm pole}} = \overline{m}_b ( 1 + c_1 \alpha_S + 
c_2 \alpha_S^2 + c_3 \alpha_S^3 )$
to eq.(\ref{totene}) and expand
$E_{\rm tot}(r; \overline{m}_b, \alpha_S(\mu), \mu)$
in $\alpha_S(\mu)$ up to ${\cal O}(\alpha_S^3)$.
Then the perturbative expansion becomes much more convergent as
well as much less dependent on the renormalization scale $\mu$
(see eqs.(\ref{seriesA})--(\ref{seriesB}) below).
The remaining renormalon is of order
$\Lambda \times (\Lambda \cdot r)^2$.
\begin{figure}[t]
\psfrag{XXX}{\footnotesize $r~[{\rm GeV}^{-1}]$}
\psfrag{YYY}{\footnotesize \hspace{-8mm} Energy~~~$[{\rm GeV}]$}
\psfrag{PowerLaw}{\footnotesize Power--law potential}
\psfrag{Log}{\footnotesize Log potential}
\psfrag{Cornell}{\footnotesize Cornell potential}
\psfrag{1161}{\footnotesize $\alpha_s(M_Z)=0.1161$}
\psfrag{1181}{\footnotesize $\alpha_s(M_Z)=0.1181$}
\psfrag{1201}{\footnotesize $\alpha_s(M_Z)=0.1201$}
\psfrag{u1s}{\footnotesize $\Upsilon(1S)$}
\psfrag{u2s}{\footnotesize $\Upsilon(2S)$}
\psfrag{u3s}{\footnotesize $\Upsilon(3S)$}
\psfrag{u4s}{\footnotesize $\Upsilon(4S)$}
\psfrag{j1s}{\footnotesize $J/\psi$}
\psfrag{j2s}{\footnotesize $\psi(2S)$}
\includegraphics[width=73mm]{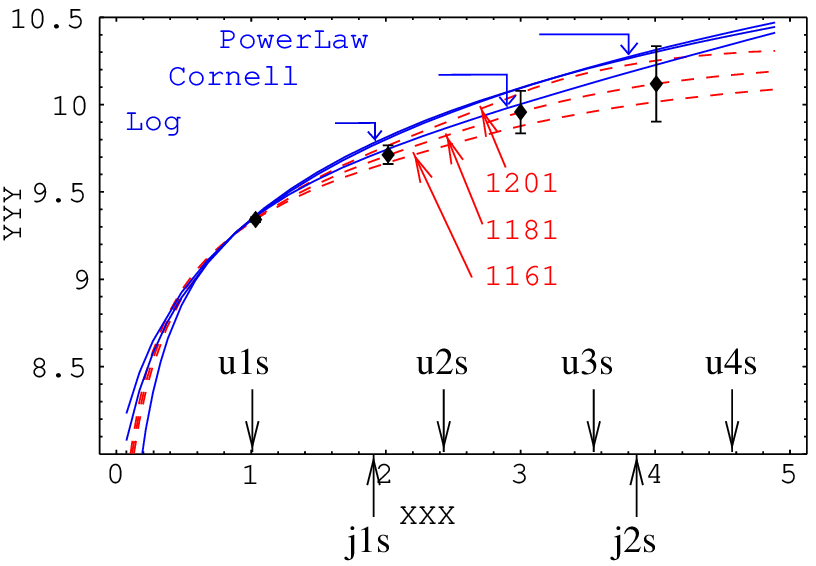}
Fig.1:~$E_{\rm tot}(r)$ and typical
phenomenological potentials.
Constants are added to make all curves coincide
at $r=1~{\rm GeV}^{-1}$.
Arrows at the bottom show the r.m.s.\ radii of the heavy
quarkonium states.
The figure is taken from \cite{RS1}.
\vspace{-4mm}
\label{comparemodels}
\end{figure}
Fig.1 shows $E_{\rm tot}(r)$ corresponding to
the present values of the strong coupling constant.
They agree well 
with typical potentials used in phenomenological model analyses
within the uncertainty estimated from the remaining 
$\Lambda^3 r^2$ renormalon
(indicated by error bars),
in the range relevant to the bottomonium and charmonium states.

Recently several comparisons 
have been made \cite{bsw,force,RS1,pineda02} 
among the perturbative predictions of
the QCD potential (in renormalon-subtracted schemes), 
the QCD potential calculated by
lattice simulations, and the phenomenological 
potentials in this region.
Combining these analyses, we find that
all these potentials agree well with one another.
It appears that the non-perturbative effects, if they exist, 
are comparable in size to the perturbative uncertainty
(apart from those in
the $r$-independent part of the potentials, which has
not been constrained by these analyses).

Phenomenological potentials in the above range may be represented
by a Coulomb-plus-linear potential, which becomes steeper than
the Coulomb potential at larger distances.
We can understand why $E_{\rm tot}(r)$ also becomes steeper than the
Coulomb potential in perturbative QCD \cite{force}.
After realizing that the ${\cal O}(\Lambda)$ renormalon should be cancelled,
it is natural to define a strong coupling constant 
$\alpha_F(\mu)$ from the interquark force:
\begin{eqnarray}
F(r) \equiv - \frac{d}{dr} \, V_{\rm QCD}(r) \equiv - C_F \, 
\frac{\alpha_F(1/r)}{r^2}.
\end{eqnarray}
Since the ${\cal O}(\Lambda)$ renormalon in $V_{\rm QCD}(r)$
is  $r$-independent, it is {\it killed} upon
differentiation.
$\alpha_F(1/r)$ grows at IR due to the running
(the first two coefficients of the $\beta$-function are universal), 
which makes $|F(r)|$
stronger than the Coulomb force at large distances.
This means that $V_{\rm QCD}(r)$, after subtraction of the renormalon,
becomes steeper than the Coulomb potential.

\subsection{Bottomonium spectrum}

The bottomonium energy levels can be computed in series expansions
in $\alpha_S$ within the boundstate theory based on perturbative QCD.
Presently the full corrections up to ${\cal O}(\alpha_S^4 m_b)$ are known
\cite{spectrum}.
It is illuminating to
compare the series expansions before and after the
cancellation of the ${\cal O}(\Lambda)$ renormalons.
For instance:\\
\underline{$\bullet \Upsilon(1S)$}:\footnote{
The parameters are taken as
$\mu = 2.49~{\rm GeV}$, $\alpha_S(\mu)=0.274$,
$\overline{m}_b=4.20~{\rm GeV}$, $m_{b,{\rm pole}}=4.97~{\rm GeV}$.
}
\bea
M_{\Upsilon(1S)} &=& 
9.94 - 0.17 - 0.20 - 0.30 ~~{\rm GeV}
\label{seriesA}
\\
&=&
8.41 + 0.84 + 0.20 + 0.013~{\rm GeV}
\eea
\underline{$\bullet \Upsilon(2S)$}:\footnote{
The parameters are taken as
$\mu = 1.09~{\rm GeV}$, $\alpha_S(\mu)=0.433$,
$\overline{m}_b=4.20~{\rm GeV}$, $m_{b,{\rm pole}}=4.97~{\rm GeV}$.
}
\bea
M_{\Upsilon(2S)} &=& 
9.94 - 0.10 - 0.19~ - 0.45 ~~{\rm GeV}
\\
&=&
8.41 + 1.46 + 0.093 + 0.009~{\rm GeV}
\label{seriesB}
\eea
The upper lines correspond to the pole-mass scheme and the lower lines
to the $\overline{\rm MS}$-mass scheme.
As can be seen, the convergence property improves dramatically 
when the $\overline{\rm MS}$ mass is used instead of the pole mass.
In this scheme, 
the perturbative series for the bottomonium energy levels
show converging behaviors up to the states with the
principal quantum number $n = 3$ \cite{physpic,bsv2}.

\begin{figure}[t]
\begin{center}
\psfrag{BSV}{BSV}\psfrag{1181}{(1181)}\psfrag{1161}{(1161)}
\psfrag{EQuigg}{Eichten--}\psfrag{EQuigg2}{Quigg}
\psfrag{Exp}{Exp}
\psfrag{3.}{\hspace{-4mm}\begin{tabular}{c}RS\\(1181)\end{tabular}}
\includegraphics[width=70mm]{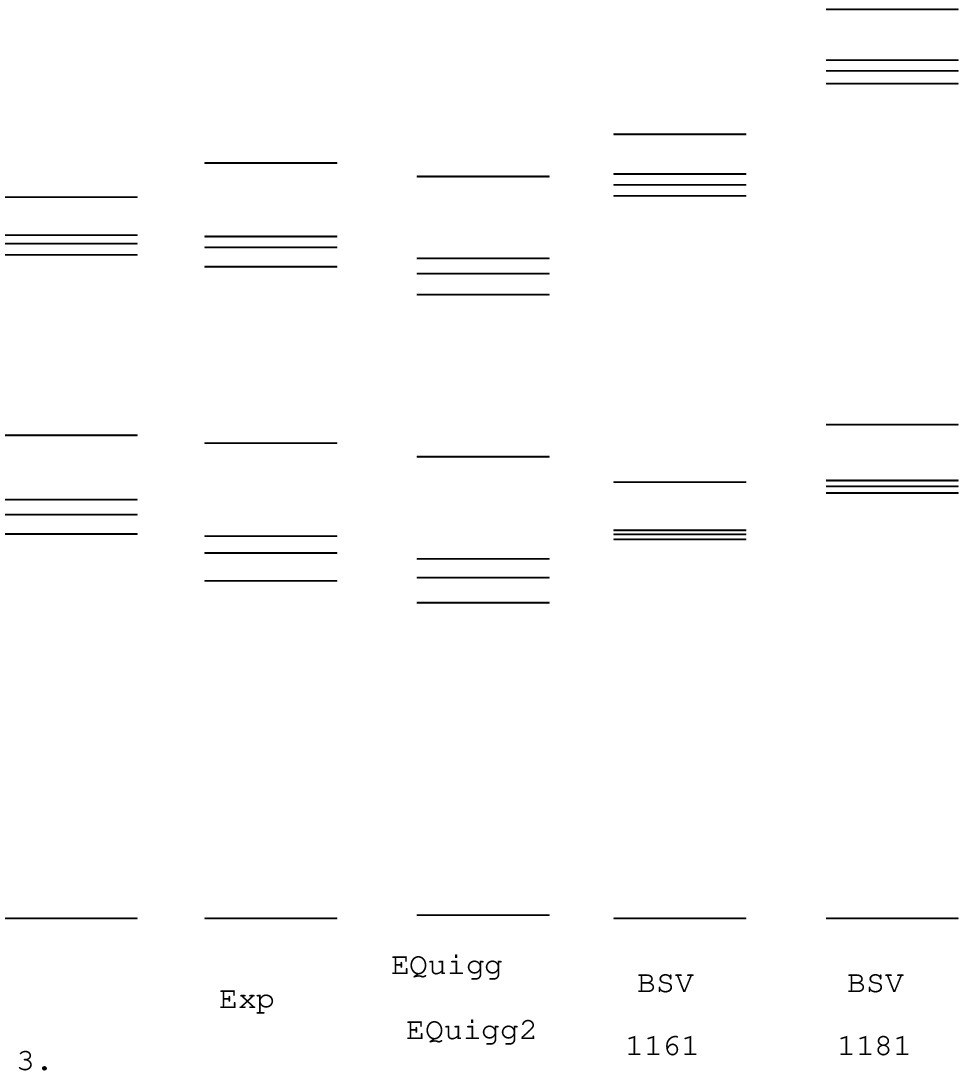}
\end{center}
\vspace{-2mm}
Fig.2:~Bottomonium spectrum up to $n=3$ as calculated in various frameworks
compared to the experimental data.
\end{figure}
Fig.2 compares the bottomonium
spectrum calculated in various
frameworks with the experimental data:
BSV(1161) and (1181) represent the fixed-order
perturbative QCD predictions up to ${\cal O}(\alpha_S^4 m_b)$
for the input $\alpha_S(M_Z)=0.1161$ and 0.1181, respectively
\cite{bsv2};
RS(1181) represents the perturbative QCD prediction 
which includes in addition part 
of the higher-order corrections, in particular the full 
${\cal O}(\alpha_S^5 m_b)$ corrections to the fine structure,
for $\alpha_S(M_Z)=0.1181$ and $\mu=3$~GeV \cite{RS2};
Eichten-Quigg represents the prediction of the
phenomenological potential model \cite{eq}.
The level of agreement between the perturbative prediction (RS)
and the experimental data approximates to that of the
Eichten-Quigg model.

\subsection{Physical picture \cite{physpic}}

When the ${\cal O}(\Lambda)$ renormalons are cancelled, 
the major part of the energy of a bottomonium state can be
written as
\bea
E \approx 2 \overline{m}_b + 
{\hbox to 18pt{
\hbox to -9pt{$\displaystyle \int$} 
\raise-21pt\hbox{$\scriptstyle R^{-1}$} 
\raise-20pt\hbox to -2pt{$\scriptstyle <$}
\raise-24pt\hbox{$\scriptstyle \sim$} 
\raise-21pt\hbox{$\scriptstyle q <\overline{m}_b$} 
}}
\frac{d^3\vec{q}}{(2\pi)^3} \, \,
C_F \frac{4\pi\alpha_S(q)}{q^2} ,
\label{app-Etot}
\eea
where $R$ represents the spacial size of the boundstate.
It shows that the 
energy is mainly composed of
(i) the $\overline{\rm MS}$ masses of $b$ and $\bar{b}$, and 
(ii) the self-energies of $b$ and $\bar{b}$ originating
from gluons whose wavelengths are shorter than the size of the boundstate
$1/\overline{m}_b \simlt \lambda \simlt R$.
As expected, contributions of IR gluons ($\lambda > R$)
have decoupled from eq.(\ref{app-Etot}).
The potential energy between $b$ and $\bar{b}$
turns out to be much smaller than the self-energies (ii).
This observation is reminiscent of the constituent quark mass 
picture for explaining the masses of light hadrons, if we regard
the energy (ii) as the difference between
(state-dependent) constituent quark masses and current quark masses.

In Fig.2 the level spacings between 
consecutive  $n$'s are almost constant,
in contrast to the Coulomb spectrum in which the spacing
decreases rapidly with $n$ as $1/n^2$. 
This is because the second term of eq.(\ref{app-Etot})
grows rapidly as the size of the boundstate increases.
Namely, the rapid growth of $\alpha_S(q)$ at $q \sim R^{-1}$
as $R$ increases
pushes up the energy levels of the excited states.
This gives a microscopic description
for how the interquark force
between a color-singlet heavy quark pair becomes strong
at large distances
(within the range where the perturbative prediction is still valid).
It is the rapid growth of the self-energies (ii) 
rather than of the potential energy
that gives the dominant effect.

\section{Cancellation of {\boldmath ${\cal O}(\Lambda^3)$} Renormalons}

Let us briefly state a historical background related to this subject.
As already noted, the energy of a static $b\bar{b}$ pair,
eq.(\ref{totene}), contains 
${\cal O}(\Lambda^3 r^2)$ renormalon, where we may
replace $r$ by the typical size $R \sim (\alpha_S m_b)^{-1}$
of the boundstate.
Within the potential-NRQCD framework, it was shown \cite{BraPineSotVai} 
that this renormalon can be absorbed
into a non-local gluon condensate, i.e.\ the perturbative 
uncertainty can be factorized and replaced by a non-perturbative
parameter of the same dimension $\sim \Lambda^3$.
On the other hand, it has been known for a long time 
\cite{vol} that the
leading non-perturbative corrections to the quarkonium 
energy levels are 
${\cal O}(\Lambda^4)$ since they are proportional
to the local gluon condensate $\langle G_{\mu \nu} G^{\mu\nu} \rangle$.
Thus, there is an apparent mismatch in the power of $\Lambda$
between the two quantities.

It has recently been shown \cite{offshell} that, if the offshellness of
$b$ and $\bar{b}$ is incorporated properly, it 
provides an additional suppression factor
of order $\Lambda/(\alpha_S^2 m_b)$ to the renormalons
in the quarkonium energy levels:
\bea
\delta E \sim \Lambda \times 
\frac{\Lambda^{\,2}}{(\alpha_S m_b)^2} \times
\framebox[12mm]{
${\displaystyle \frac{\Lambda}{\alpha_S^2 m_b}}$
} .
\eea
Hence, the dimension of the perturbative uncertainty becomes
the same as that of the leading non-perturbative corrections 
(including the powers of $\alpha_S$ and
$m_b$ in the denominator).
Moreover, convergence of the perturbative series
improves by this effect.

Intuitively the suppression mechanism may be understood as follows.
Large offshellness of $b$ and $\bar{b}$ corresponds to 
short rescattering time of $b$ and $\bar{b}$ inside the boundstate.
If the rescattering time
$\Delta t \sim (\alpha_S^2 m_b)^{-1}$ is shorter than the
hadronization time $\sim \Lambda^{-1}$,
$b$ and $\bar{b}$ will get distorted before 
IR gluons surround them and an energy 
is accumulated by antiscreening effects.
Thus, we expect the offshellness to act
as an infrared cutoff to the effects induced by 
gluons' timelike propagation.

In the (leading) kinematical configuration relevant to 
formation of the bottomonium states, gluons exchanged
between $b$ and $\bar{b}$ have momenta $q^\mu$ where
$ |q^0| \ll |\vec{q}|$.
Therefore, in the conventional approach we take the
instantaneous gluon-exchange as the leading-order
and incorporate perturbations to it.
On the other hand, a different kinematical region, 
$-q^2 \sim \Lambda^2$,
contributes to renormalons.
Because of this, in order to estimate renormalons
in the bottomonium spectrum accurately, we need to make an approximation 
valid in both of the above kinematical regions.

An analysis following this observation reveals 
that there is a rich mathematical
structure underlying the renormalon suppression effect.
For instance,
the expansion about the instantaneous potential changes the
analyticity of the Borel transform $\widetilde{E}[u]$ of the
bottomonium energy level.
IR renormalons, defined as poles 
of $\widetilde{E}[u]$ in the $u$-plane, 
are located at $u = 2,3,4,\dots$.
But once the expansion is made, poles are generated at
$u=\frac{3}{2},\frac{5}{2},\frac{7}{2},\dots$, 
while at the leading-order of the
expansion, the original poles at $u = 2,3,4,\dots$ disappear.
For example, at $u=\frac{3}{2}$,
there is a cancellation of poles in the combination
\bea
\widetilde{E}[u] \sim
\frac{\Delta^{3-2u}-1}{u-{3}/{2}} , 
~~~~~
\mbox{where}~~
\Delta = {\textstyle \frac{1}{2}}{C_F \alpha_S}.
\eea
$\Delta$ represents the offshellness or non-instantaneity
of the potential.
However, if $\Delta$ is considered to be small and an
expansion
$ \Delta^{3-2u} = \Delta^3 \times ( 1 - 2u\log\Delta + \dots )$
is performed, there remains  
an uncancelled pole at $u=\frac{3}{2}$.
Hence, the perturbative uncertainty becomes larger by this expansion.

\section{Conclusions}

There has been important progress in the theory
of heavy quarkonium states, which extended the predictive power 
of perturbative QCD beyond what could be achieved before.
Perturbative predictions 
can be made accurate once IR degrees of freedom are eliminated
properly from the calculations.
Renormalon dominance picture suggests that uncertainties of 
the perturbative prediction for the bottomonium spectrum are of moderate 
size, of 
${\cal O}(\Lambda^3/(\alpha_S^2 m_b^2))$ if the short-distance mass
is used, of 
${\cal O}(\Lambda^4/(\alpha_S^4 m_b^3))$ if the offshell effects
are further incorporated.
Evidence has been found for the static QCD potential and the
bottomonium spectrum, which supports 
the hypothesis that the
perturbative predictions agree with the full QCD predictions
within the estimated moderate uncertainties.

Based on the perturbative QCD predictions, we obtained new microscopic
pictures on the composition of the heavy quarkonium spectrum and on the
behavior of the interquark force in the intermediate distance region.
Furthermore, the quark offshell effect as an IR cutoff in 
timelike processes was recognized, whose nature is
essentially non-local in time.

So far, most important applications of the progress
are the precise determinations of the $\overline{\rm MS}$ masses of
$b$ and $c$ quarks: e.g.\
$\overline{m}_b = 4190 \pm 32$~MeV,
$\overline{m}_c = 1237 \pm 60$~MeV have been obtained 
based on the above hypothesis that non-perturbative corrections
can be absorbed into perturbative uncertainties \cite{bsv2}.

\section*{Acknowledgements}
The works reported in this article
are based on the collaborations with 
N.~Brambilla, A.~Vairo, Y.~Kiyo and S.~Recksiegel.
The author is grateful to all of them for very fruitful discussion.

\end{document}